\documentclass[a4paper,10pt,twocolumn,aps,pra,superscriptaddress,nofootinbib,longbibliography,showpacs,tightenlines]{revtex4-1}

\usepackage[utf8]{inputenc}
\usepackage[T1]{fontenc}

\usepackage{graphicx}
\usepackage[dvipsnames]{xcolor}
\graphicspath{{./fig/}{./tikz/}}

\usepackage{amssymb,amsmath,braket}
\usepackage{siunitx}

\usepackage{booktabs}

\newcommand{\ham}[1]{\mathcal{#1}}
\newcommand\mechpos{x}
\newcommand\xzpf{\ensuremath{x\s{zp}}}
\newcommand{\MM}[1]{M$_{#1}$}
\newcommand\abs[1]{\left| #1 \right|}
\newcommand{\s}[1]{_\mathrm{#1}}
\newcommand*{\up}[1]{\ensuremath{^\text{#1}}}
\newcommand*{\avg}[1]{\left\langle #1 \right\rangle}
\newcommand*{\symm}[2]{\avg{ #1 \circ #2}}
\newcommand{\Mode}[1]{\mathcal{#1}}
\newcommand{\ci}{\mathcal I}

\usepackage{verbatim}

\definecolor{darkblue}{rgb}{0,0,0.7}
\definecolor{darkred}{rgb}{0.7,0,0}
\definecolor{darkgreen}{rgb}{0,0.5,0}
\definecolor{added}{rgb}{0.8,0.4,0}

\usepackage[unicode, colorlinks, citecolor=darkblue, linkcolor=darkred, urlcolor=blue]{hyperref}

\begin{document}
\title{Pulsed quantum interaction between two distant mechanical oscillators}
\author{Nikita \surname{Vostrosablin}}
\email{nikita.vostrosablin@upol.cz}
\author{Andrey A. \surname{Rakhubovsky}}
\author{Radim \surname{Filip}}

\affiliation{Department of Optics, Palack{\'y} University, 17. Listopadu 12, 771 46 Olomouc, Czech Republic }

\pacs{42.50.Wk,42.50.Dv,42.50.Ex}

\begin{abstract}
	Feasible setup for pulsed quantum non-demolition interaction between two distant mechanical oscillators through optical or microwave mediator is proposed. The proposal uses homodyne measurement of the mediator and feedforward control of the mechanical oscillators to reach the interaction. To verify quantum nature of the interaction, we investigate the Gaussian entanglement generated in the mechanical modes. We evaluate it under influence of mechanical bath and propagation loss for the mediator and propose ways to optimize the interaction. Finally, both currently available optomechanical and electromechanical platforms are numerically analyzed. The analysis shows that implementation is already feasible with current technology.
\end{abstract}

\maketitle

\section{Introduction} 
\label{sec:introduction_and_summary}

Quantum optomechanics and electromechanics connecting light and microwaves with mechanical motion at the quantum level is an emerging field of quantum physics and technology~\cite{aspelmeyer_cavity_2014, Aspelmeyer2014, Bowen2015}. Recently, Gaussian quantum entanglement between mechanical oscillator and microwave field~\cite{Palomaki2013}, and nonclassical photon-phonon correlation of mechanical membrane and optical pulse~\cite{riedinger_non-classical_2016} have been experimentally demonstrated. Both experiments used modern pulsed optomechanics~\cite{hofer_quantum_2011,vanner_pulsed_2011,vanner_cooling-by-measurement_2013,Liao2011_1,Liao2011_2}. They open new possibilities to experimentally connect other physical platforms with mechanical oscillator, like continuous-variable cold atom ensembles~\cite{hammerer_establishing_2009, Camerer2011, Jockel2015}, and further many discrete systems like individual atoms~\cite{hammerer_establishing_2009, Vogell2015}, superconducting qubits~\cite{OConnell2010, Pirkkalainen2013}, solid-state systems~\cite{Arcizet2011, kolkowitz_coherent_2012, Teissier2014, ovartchaiyapong_dynamic_2014} and semiconductor systems~\cite{Yeo2014,Montinaro2014}.  Together with these interesting and challenging interdisciplinary experiments, state-of-the-art of laboratory techniques could currently allow to let interact two mechanical oscillators mediated by light or microwave field. It is another interesting step forward, two similar mechanical oscillators coupled at quantum level have not been demonstrated yet. It can be very stimulating especially because the connection between two mechanical systems can physically connect quantum optomechanics to classical thermodynamics. If two similar quantum mechanical oscillators will be interfaced by the quantum version of the coupling typically used in classical mechanics, they can naturally generate entanglement. It is a simple witness that they were coupled quantum mechanically. Additionally, the mechanical-mechanical interaction can be quantum non-demolition type, which is required for basic continuous-variable quantum gate~\cite{Bartlett2002} , useful for its specific features, for both gate-based~\cite{Yoshikawa2008} and cluster-state-based~\cite{Miwa2010} quantum computing. Recently, the nonlocal optical QND gate was demonstrated~\cite{Yokoyama2014} following the theoretical proposals in~\cite{filip_continuous-variable_2004, FilipMarek2005}. Such the QND coupling was already broadly exploited between two atomic ensembles~\cite{Hammerer2010}. It is therefore much more important for the future to achieve such the well-defined quantum interaction of mechanical oscillators, not only the generation of entangled state of two mechanical systems.

Generation of entanglement between two mechanical systems have been already proposed in three different configurations. In the first type of proposed setups, two mechanical oscillators have been placed in a single optical cavity~\cite{Mancini2002, Hartmann2008,Xuereb2012, Woolley2013, Seok2013, Tan2013, Woolley2014, Flayac2014, Yang2015}. In this case, the continuous generation of steady-state entanglement appears because the mechanical oscillators interact with join optical intra-cavity field. This configuration has been extensively used to discuss continuous-time quantum synchronization~\cite{Mari2013, Ying2014, Weiss2016}. In the second kind of proposals, two entangled beams of light were used to entangle two mechanical systems without necessity to measure them~\cite{Zhang2003,Pinard2005, Mazzola2011}. In the third kind of proposed setups, two continuous-wave beams of light, leaving two continuously pumped optomechanical cavities, are jointly detected in Bell measurement and photocurrent is used to correct the mechanical oscillators~\cite{Pirandola2006, Borkje2011, Abdi2012} . These schemes can generate entanglement at a distance, however, it is very limited because of instabilities in the blue-detuned continuous-wave regime. Advanced time-continuous quantum measurement and control has been suggested to prepare mechanical entanglement~\cite{Hofer2015}. Recently, theoretical investigation of optomechanical crystals has offered many other ways how to obtain mechanical entanglement~\cite{Schmidt2012, Flayac2015}. Our goal is to propose currently feasible scheme with potential to use power of quantum optics tools to complement recent experimental test of coupled quantized mechanical oscillations of trapped ions~\cite{Brown2011}.

In this paper, we propose currently feasible way to build basic {\em pulsed} quantum non-demolition (QND) interaction between two mechanical oscillators at a distance, connected by light or microwave field. The scheme is depicted in Fig.~1. Using homodyne detection of light or microwave field and feedforward control, means of both mechanical oscillators precisely follow the QND interaction.
To generate significant entanglement of mechanical oscillators, coherent light is sufficient, and the entanglement can be very well estimated when intra-cavity field can be adiabatically eliminated. On the other hand, squeezed light is advantageous to approach ideal QND interaction between two mechanical systems. Feasible squeezing of light is capable to enhance entangling power of the QND interaction. However, for larger optomechanical coupling strength and larger squeezing, non-adiabatic methods taking the intracavity field fully into account are required. Importantly, the non-adiabatic calculations predict a decrease of the entanglement power for larger squeezing. It is due to presence of the intra-cavity field and the squeezing has to be therefore optimized to get maximum of entangling power. We prove sufficient stability of the QND interaction under influence of mechanical bath and transmission loss between two separated cavities. Finally, we verified that it is feasible to build the mechanical QND interaction for both current optomechanical~\cite{meenehan_pulsed_2015} and electromechanical~\cite{ockeloen-korppi_low-noise_2016} setups.

The paper is organized in the following way. We begin by mathematical definition of quantum-nondemolition interaction and principal description of the experimental setup.  First, in Sec.~\ref{sec:principal_performance} we carry out a simple principal analysis of the physics of  the setup. To do so we start from a brief derivation of equations of motion for an optomechanical system in Sec.~\ref{sec:optomechanical_quantum_non_demolition_interaction} and solve those in Sec.~\ref{sec:adiabatic_regime} ignoring for a while the decoherence and eliminating intracavity modes. We quantify the interaction between the mechanical modes analyzing the transfer of first moments of quadratures, and for a figure of merit of the strength of the interaction we employ the entanglement between the modes. We use logarithmic negativity~\cite{serafini_symplectic_2004} as a measure of entanglement. We show principal possibility of the protocol performance and derive the simplest conditions on the experimental parameters.

Second, in Sec.~\ref{sec:robustness_to_imperfections} we perform a full numerical analysis of the system allowing for the imperfections. Those include impact of the intracavity modes that mediate the interaction between the travelling light pulse and the mechanical modes and the thermal bath causing decoherence of the latter. We as well investigate the impact of the optical loss between the cavities. We show that with currently available parameters the protocol can establish a QND interface between the two distant mechanical modes.

\section{Setup for pulsed QND interaction between mechanical oscillator} 

\begin{figure}[t]
	\includegraphics[width=.99\linewidth]{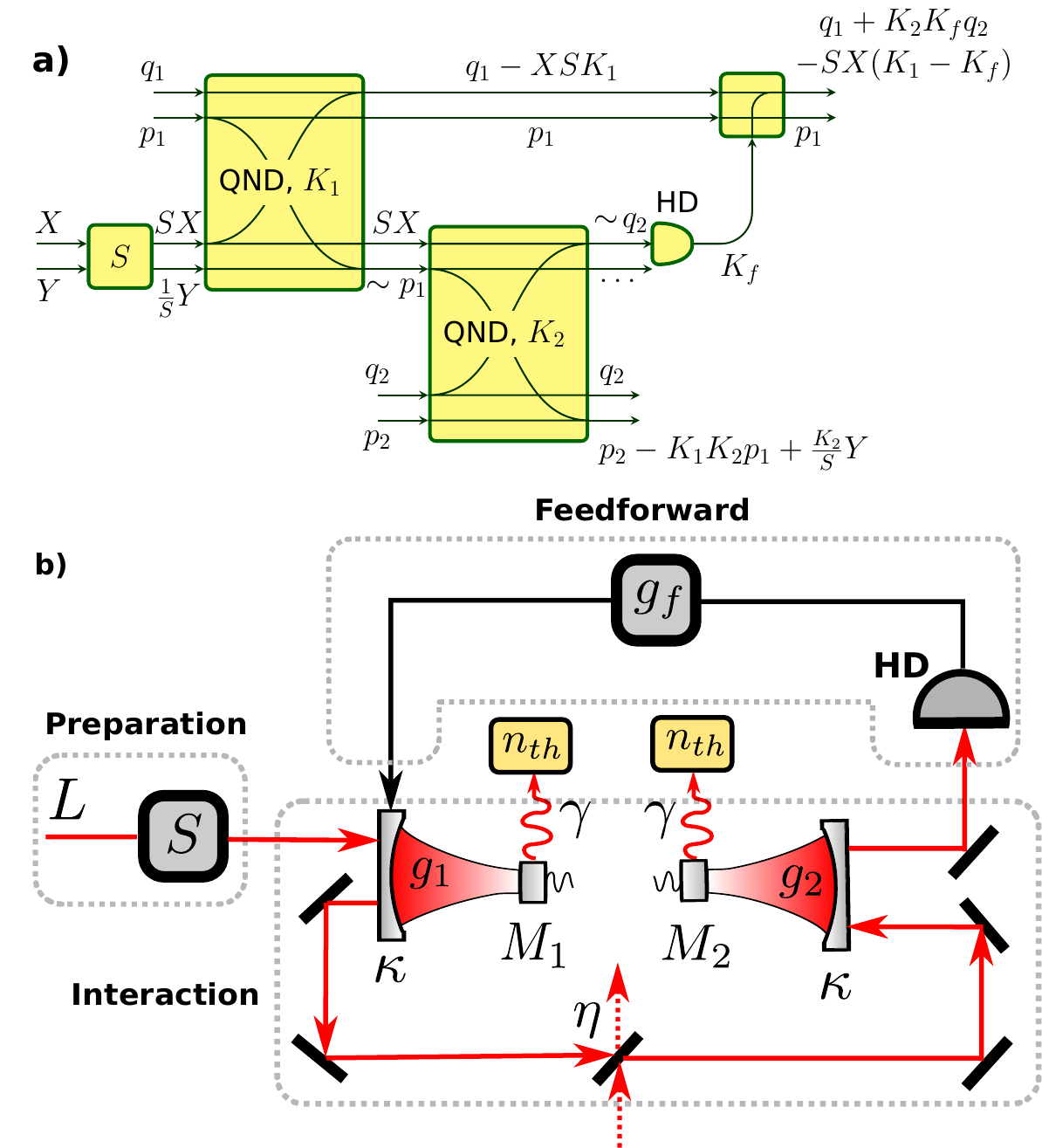}
	\caption{\label{fig:protocol}
		The protocol of QND interaction between two mechanical modes. a) Simplified scheme,  $S$~--- squeezing operation, HD~--- homodyne detector. b) possible experimental implementation with the imperfections: $\eta$~--- optical losses between the cavities, $n\s{th}$~--- thermal mechanical environment.
	}
\end{figure}

In this paper, we propose a feasible way of implementation of quantum non-demolition (QND) interaction between mechanical modes of two distant optomechanical cavities.  The QND interaction of two harmonic oscillators may be described by Hamiltonian $\ham H_{\s{int}} = \hbar g Q_1 Q_2$ with $Q_{1,2}$ being  the position or momentum of the corresponding oscillator and $g$, interaction strength. After the interaction  both the variables $Q_1$ and $Q_2$ remain unperturbed (not demolished) whereas the  complementary ones to $Q_1$ ($Q_2$) become linearly displaced by a value proportional to $gQ_2$ ($gQ_1$).  The nondemolition interaction has been demonstrated in a few electromechanical experiments recently~\cite{wollman_quantum_2015,pirkkalainen_squeezing_2015}.

The proposed scheme is presented in the Fig.~\ref{fig:protocol}.
It is the simplest setup for generation of QND coupling between two mechanical systems. It is basically a serial scheme which does not require multiple pass of optical pulse through single optomechanical cavity. Moreover, it exploits advantage of squeezed light, homodyne detection, which are very efficient resources of quantum optics. The feedforward correction on mechanics can be done simply at any time by classical pulse of laser light.
The  modes of two mechanical oscillators \MM1 and \MM2 interact by turns with an  optical (or microwave) pulse $L$ via opto(electro-)mechanical coupling. The pulse is then detected and the result of the detection is used to  linearly displace the mechanical mode of the first cavity  (if needed, in the second one is displaced as well).  In principle this feedforward is not necessary to achieve entanglement, as the latter could be created contional on the results of the detection.  A similar method was used recently for conditional state preparation in optomechanics~\cite{Wieczorek2015}.

The  optical pulse can be prepared in a squeezed intensive coherent state and sent into the optomechanical cavity.
The latter in essence comprises an optical mode coupled via radiation pressure to a mechanical harmonic oscillator~\cite{law_interaction_1995}.
We follow the standard approach~\cite{braginsky_quantum_1980,clerk_back-action_2008,aspelmeyer_cavity_2014} and assume that the optical pulse is displaced with a strong classical component that is modulated at mechanical frequency. This ensures that the effective interaction within the cavity is the non-demolition type.

The QND interaction allows a partial exchange of the variables between the mechanical mode \MM1 and the travelling pulse (see Fig.~\ref{fig:protocol}~(a)). The latter is then redirected to the second cavity  with mechanical oscillator \MM2, which we assume to be identical to the \MM1. The QND interaction within the second cavity allows to transmit  a variable of the mode \MM1 carried by the pulse to the mode \MM2 and in turn to transmit  a variable of the mode \MM2 to the light. The pulse is then detected and the result of detection is used to displace the mode \MM1 to complete transfer of the \MM2  variable. A proper strong presqueezing of the light pulse  and the feed-forward correction allow to eliminate  all variables from the final transformation of the mechanical modes that  consequently approach an ideal QND interaction between them.

\section{Performance of setup for QND interaction} 
\label{sec:principal_performance}
\label{section1}

\subsection{Optomechanical quantum non-demolition interaction} 
\label{sec:optomechanical_quantum_non_demolition_interaction}

Let us first consider a single optomechanical cavity that in essence embodies an optical mode and a mechanical one.  The two modes are coupled by radiation pressure with the Hamiltonian~\cite{law_interaction_1995} $H\s{rp} = - \hbar g_0 n\s{cav} \mechpos/\xzpf$, where $n\s{cav}$ stands for intracavity photon number, $\mechpos$, mechanical displacement from equilibrium, $g_0$, so-called single-photon coupling strength. The mechanical zero-point fluctuation amplitude, denoted by \xzpf, for a mechanical oscillator with mass $m$ and eigen frequency $\omega_m$ equals $\xzpf = \sqrt{ \hbar / 2 m \omega_m}$.

In order to enhance the radiation pressure coupling, strong coherent field is used as the pump. This allows to linearise the dynamics around a steady classical state and solve for quantum corrections.   Moreover, we assume this strong classical field to be resonant with the cavity and modulated at the frequency of the mechanical oscillator~\cite{braginsky_quantum_1980}. In this case if the mechanical frequency $\omega_m$ exceeds all other characteristic frequencies of the system, one can perform averaging to get rid of the terms at $2 \omega_m$ (i.e., adopt the Rotating Wave Approximation, RWA) to obtain the non-demolition coupling.  The latter condition is usually equivalent to the requirement that the optical decay rate $\kappa$ of the cavity be smaller with respect to $\omega_m$.,  known as resolved-sideband regime.

After the linearization and averaging out the rapid oscillating terms we arrive to the QND coupling within the optomechanical cavity with Hamiltonian that reads (depending on the phase of the pump)
\begin{equation}
	\ham H = \hbar g X p
	\quad \text{ or }
	\quad
	\ham H = \hbar g Y q,
\end{equation}
where $g = g_0 \sqrt{ \avg{n\s{cav}}}$ is the enhanced optomechanical coupling strength, $X$ and $Y$, and $q$ and $p$ are quadratures of, respectively, the optical and mechanical modes which obey usual commutation relations ($[X,Y] = i; \ [q,p] = i$). The mechanical displacement $\mechpos$ can be expressed in terms of quadratures as $\mechpos  / \xzpf = q  \cos \omega_m t + p  \sin \omega_m t$ and a similar expression holds for the optical quadratures.

The counter-rotating terms at $2 \omega_m$ could provide additional back-action. In Appendix~\ref{sec:rotating_wave_approximation} we analyze this back-action and prove that for typical experimental parameters it is sufficient to consider the system within RWA.

To describe the interaction of the propagating light pulse with the optomechanical cavity we complement the Hamiltonian of the optomechanical interaction $\ham H_1 = -\hbar g_1 X_1 p_1$ with input-output relations~\cite{walls_quantum_2007} (henceforth we denote with index ``1'' or ``2'' quantities corresponding to the respective cavity). The system is thus described by the following set of equations:
\begin{gather}
	\notag
	\begin{aligned}
		\dot{q}_1 & = - \tfrac \gamma 2 q_1 -g_1 X_1 + \xi_{q1},
				  &
		\dot{X}_1 & = - \kappa X_1 + \sqrt{2 \kappa} X\up{in},
		\\
		\dot{p}_1 & = - \tfrac \gamma 2 p_1 + \xi_{p1},
				  &
		\dot{Y}_1 & = - \kappa Y_1 + \sqrt{2 \kappa} Y\up{in} + g_1 p_1
	\end{aligned}
	\\
	\label{eq:1stqnd}
	Q\up{out} = \sqrt{ 2 \kappa } Q - Q\up{in},
	\quad
	Q = X,Y.
\end{gather}
Here $X\up{in},Y\up{in}$ are the quadratures of the pulse with commutator $[X\up{in} , Y\up{in} (t')] = i \delta (t - t')$, $\xi_{{q,p}}$ are the quadratures of mechanical noise. $\kappa$ and $\gamma$ are respectively optical and viscous mechanical damping coefficients.

\begin{figure}[t]
	\centering
	\includegraphics[width=.99\linewidth]{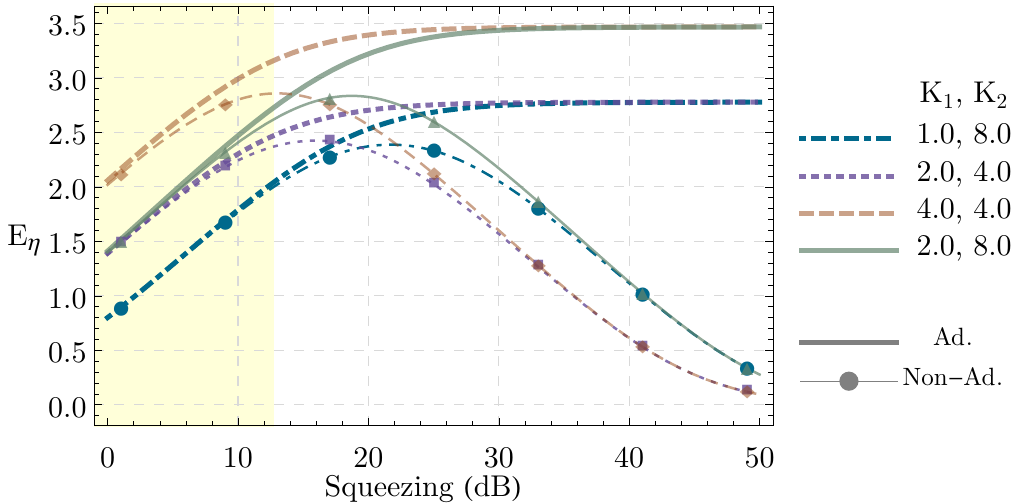}
	\caption{
	Entanglement between the two mechanical modes as a function of optical presqueezing. Thick lines correspond to adiabatic solution; thin lines with markers to solution with cavity mode. Different colors and dashings are used for different ratio of the gains $K_1$ and $K_2$. Losses are absent: $\eta =1$. Highlighted is the region of squeezing magnitudes not exceeding the value of $\SI{12.7}{\decibel}$ reported in Ref.~\cite{eberle_quantum_2010}.
	}
	\label{fig:coupling}
\end{figure}

\subsection{Adiabatic regime} 
\label{sec:adiabatic_regime}

As a first approximation we consider the system in adiabatic regime.
Given that optical decay rate exceeds the other rates in~\eqref{eq:1stqnd} (which is typically the case in experiment), one can assume that the optical mode reacts to any changes instantaneously, which is equivalent to putting $\dot X = \dot Y = 0$ in Eqs~\eqref{eq:1stqnd}.
Formally this corresponds to replacement of all the functions of time with their own versions averaged over the interval with duration $\tau_*$ such that $\kappa \gg 1/\tau_* \gg \gamma,g$.

Lastly, in this section we leave out the mechanical decoherence, setting $\gamma = 0,\ \xi_{q1} = \xi_{p1} =0$.

With these assumptions the solution of Eqs.~\eqref{eq:1stqnd} reads
\begin{align*}
	q_1(\tau) & = q_1(0) - S K_1 \Mode X\up{in}, &  \Mode X\up{out}_1 & = S \Mode X\up{in}, \\
	p_1(\tau) & = p_1(0) , &  \Mode Y\up{out}_1 & = \frac{1}{S} \Mode Y\up{in} + K_1 p_1(0).
\end{align*}
We have  introduced the squeezing magnitude $S$ and the effective interaction strength $K_1 = g_1 \sqrt{2 \tau/\kappa}$. We also have defined the input and output quadratures of the cavity as
\begin{equation}
	\notag
	\Mode Q\up{k} = \frac{ 1 }{ \sqrt \tau } \int_0^\tau Q\up{k} (s) ds,
	\quad
	Q = X,Y,
	\quad
	\text{k}=\text{in,out}.
\end{equation}
The quadratures are normalized to obey $[\Mode X\up{k}, \Mode Y\up{k}] = i$.

The output field from the first cavity is then delivered to the input of the second one through a purely lossy channel that performs an admixture of vacuum to the signal, therefore
\begin{equation}
	\notag
	Q\up{in}_2 = \sqrt{ \eta } Q\up{out}_1 + \sqrt{ 1 - \eta} Q\s{ls},
	\quad
	Q = X,Y.
\end{equation}
Here $Q\s{ls}$ are the quadratures of vacuum mode.

The optomechanical interaction within the second cavity is described by the Hamiltonian~$\ham H_2 =  \hbar g_2 Y_2 q_2$ and starts at time $t = \tau$. One can obtain the input-output relations for the second cavity in a similar fashion.  For simplicity we assume the parameters of the second cavity (except the coupling $g_2$) to replicate the  parameters of the first one.

The optical output quadrature $\Mode X\up{out}_2$ is measured and the position of the mechanical mode of the first cavity is displaced so that the final value equals $q_1 = q_1 (\tau) + K_f \Mode X\up{out} _2$.

\begin{align}
	\label{eq:interloss}
	q_1 & = q_1(0) + K_2 K_f q_2(\tau)
	\\ \notag
	& - S \Mode X\up{in}\left(K_1 - K_f \sqrt{\eta} \right)
	+ \Mode X\s{ls} K_f \sqrt{1 - \eta},  \\ \notag
	p_1 & = p_1(0), \\ \notag
	q_2 & = q_2(\tau),  \\ \notag
	p_2 & = p_2(\tau) - K_1 K_2 p_1(0) \sqrt{\eta}
	\\ \notag
	& - \Mode Y\up{in} \frac{K_2   \sqrt{\eta}}{S}
	-  K_2 \sqrt{1 - \eta}\Mode Y\s{ls}.
\end{align}
Similarly, we have introduced $K_{2} = g_{2} \sqrt{\frac{2 \tau}{\kappa}}$ here.

To approach the ideal QND interaction of the two mechanical modes with Hamiltonian $\ham H\s{QND} = \hbar K_1 K_2 \tau^{-1} p_1 q_2$ one needs to fulfill a few conditions. First, ensure low loss ($\eta \to 1$) to get rid of the noisy mode $Q\s{ls}$. Second, pick a proper feed-forward gain $K_f = K_1 / \sqrt{ \eta}$ and provide high squeezing $S \gg 1$ to suppress the optical mode $\Mode Q\up{in}$.

To quantify the strength of the interaction we estimate the entanglement between the two mechanical modes, namely the logarithmic negativity~\cite{serafini_symplectic_2004} (see Appendix for details).

In the lossless case the optimal value of squeezing yielding maximum of entanglement is given by $S = \abs{ K_2 / ( K_1 - K_f )}$. Therefore for the feedforward $K_f = K_1$ the entanglement increases with squeezing infinitely.  In the limit of moderately strong coupling ($K_{1,2} \gtrsim 1$) the following approximation holds:
\begin{align}
	\label{eq:symplectic}
	E_\eta \approx - \ln \frac{ 1 }{ 2 K_1 K_2 } \sqrt{ 1 + \frac{ K_2^2 }{ S^2 }}.
\end{align}
From this expression follows that although increase of both $S$ and $K_{1,2}$ leads to stronger entanglement, it is more efficient to increase $K_1$. This can be seen from the latter equation in~\eqref{eq:interloss}, where the noisy mediator quadrature $\Mode Y\up{in}$ enters with a multiplier~$\propto K_2$.

The LN for this simple model is presented as a function of the presqueezing $S$ in Fig.~\ref{fig:coupling} (solid lines). The parameters used for simulation are
$\kappa/ 2 \pi = \SI{221.5}{\mega\hertz}$, $\gamma / 2 \pi = \SI{328}{\hertz}$, $\tau = \SI{4.5}{\micro\second}$ that correspond to a recent optomechanical experiment~\cite{meenehan_pulsed_2015} with increased pulse duration~$\tau$.

From the Fig.~\ref{fig:coupling} is is clear that for low squeezing the LN is mostly defined by the interaction strength $K_1$ in the first cavity as it follows from~\eqref{eq:symplectic}. In the limit of high squeezing the LN saturates to the value that is defined by the product of gains $K_1 K_2$, again in agreement with~\eqref{eq:symplectic}.


\section{Robustness to imperfections} 
\label{sec:robustness_to_imperfections}

There are two sources of hindrance that we left out for the previous section. First is the intracavity modes that mediate interaction between the propagating pulse and the mechanical modes of interest. As well the intracavity modes produce unwanted memory effects that disturb the desired QND interaction. Second is the interaction of mechanical modes with the thermal environment.

In this section we first study these two sources independently and finally provide a full solution taking both into account simultaneously.
\subsection{Impact of the intracavity modes} 
\label{sec:impact_of_the_intracavity_modes}

To consider the effect of the intracavity modes on the QND interface, we solve the set of dynamical equations~\eqref{eq:1stqnd} without the mechanical decoherence ($\gamma = 0$, $\xi_{{q,p}} = 0$). The solution reads (for compactness we write the solution for the lossless case, $\eta =1$)
\begin{widetext}
	\begin{align}
		\label{eq:BAE}
		q_1 = & q_1(0) + q_2(\tau) K_2 K_f \left( 1 - \frac{1 - e^{-\kappa \tau}}{\kappa \tau} \right) - S (K_1 - K_f )\frac{1}{\sqrt \tau} \int_0^{\tau}  X\up{in}_1(s) ds
		\\ \notag
		& + S K_1 \int_0^{\tau} X\up{in}_1(s)\left(e^{-\kappa(\tau - s)}\left[1 - 4\kappa (\tau - s)\frac{K_f}{K_1} \right] \right) ds
		+ X_1(0)\left( \frac{ 2 g_f}{\kappa} \left[ (1 - e^{-\kappa \tau}) - 2 \kappa \tau e^{-\kappa \tau} \right] - \frac{g_1}{\kappa} \left[ 1 - e^{-\kappa \tau} \right] \right)
		\\ \notag
		& + X_2(0) \frac{2 g_f}{\kappa} \left( 1 - e^{-\kappa \tau} \right),
		\\ \notag
		p_1 = & p_1(0),
		\\ \notag
		q_2 = & q_2(\tau),
		\\ \notag
		p_2 = & p_2(\tau)
		- p_1(0) K_1 K_2 \left( 1 + e^{-\kappa \tau} - \frac{2}{\kappa \tau}(1 - e^{-\kappa \tau})  \right)
		- \frac{K_2}{S} \frac{1}{\sqrt{\tau}} \int_0^{\tau} \left( 1 -  e^{-\kappa(\tau - s)}[2 \kappa ( \tau - s) + 1] \right) Y_1\up{in}(s) ds
		\\ \notag
		& - Y_1(0) \frac{ 2 g_2}{ \kappa } \left( 1 - e^{-\kappa \tau} (1 + \kappa \tau) \right)
		- Y_2(0) \frac{g_2}{\kappa}(1 - e^{-\kappa \tau}),
	\end{align}
\end{widetext}
where we defined $g_f \equiv K_f \sqrt{ \kappa / 2 \tau }$.

These equations deviate from the idealized set~\eqref{eq:interloss} by presence of the initial intracavity quadratures $Q_{1,2} (0)$. As well the pulse quadratures $Q\up{in}$ can no longer be eliminated completely by a proper choice of $K_f$ and high squeezing $S$. Moreover, in this case high squeezing apmlifies the noisy summand with $X\up{in}$ degrading the interface.  The impact of this summand can be reduced by redefining the temporal mode of the output pulse (applying optimal time filter at the detection).  This, however, cannot cancel the noisy summand completely as the optical quadratures that are written during the first pass ($X\up{in}$) and second pass ($Y\up{in}$) are distorted in different manner, see Eq.~\eqref{eq:BAE}.

From the Eqs.~\eqref{eq:BAE} follows that in the limit $\kappa \gg g_{1,2,f}$ and $\kappa \tau \gg 1$ these equations reduce to the pure QND transformations~\eqref{eq:interloss}. Furthermore, from the first equation it follows that the effect of the unwanted summand $\propto S X\up{in}$ can be reduced by decreasing $K_1$. This is illustrated in Fig.~\ref{fig:coupling} where we plot the LN for solution including the cavity modes as a function of squeezing for different couplings. At high squeezing the full solution deviates from the adiabatic one, however the curves with lower $K_1$ show this deviation at higher squeezing than the curves with higher $K_1$.

The proper choice of the coupling thus allows to approach the performance of the idealized adiabatic regime.  Note that in order to increase the LN it is more efficient to increase $K_1$ than $K_2$.  To increase the LN staying close to the preferred adiabatic regime (and therefore a pure QND interface between the mechanical modes) on the contrary it is preferable to increase $K_2$.

\subsection{Mechanical thermal bath} 
\label{sec:mechanical_thermal_bath}

\begin{figure}[t]
	\centering
	\includegraphics[width=.99\linewidth]{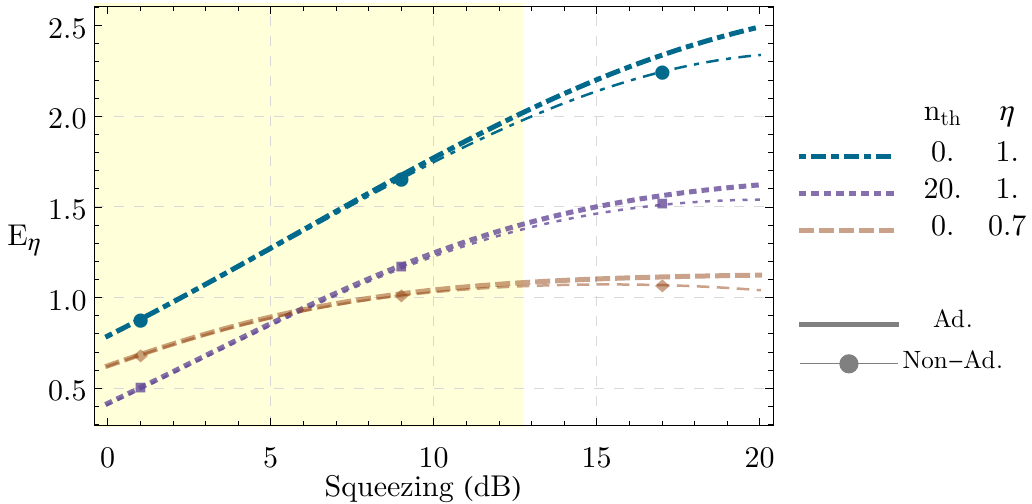}
	\caption{Entanglement as a function of squeezing in presence of mechanical bath with mean number of phonons $n\s{th}$ and optical losses with transmittivity $\eta$. The optomechanical gains equal $K_1 = 1,\ K_2 = 8$, same as for the blue dot-dashed line in Fig.~\ref{fig:coupling}}
	\label{fig:loss_noise}
\end{figure}

Finally we consider the system in presence of the thermal mechanical environment.

We assume that each of the mechanical modes is coupled at rate~$\gamma$ to its own environment that is in a thermal state with occupation $n\s{th}$ (see Fig.~\ref{fig:protocol}). The coupling for both modes takes place during the interaction with the pulse.  Moreover, the first mode remains coupled to the environment during the interaction of the second system with the pulse. Before the interaction with the pulse the mechanical modes are in the ground state (the possibility to precool mechanical oscillator close to the ground state has been demonstrated for a number of setups~\cite{chan_laser_2011,teufel_sideband_2011,riedinger_non-classical_2016}).

The thermal bath is represented in the equations~\eqref{eq:1stqnd} by Langevin force quadratures $\xi\s{q,p}$.  These quadratures are assumed Markovian so that
\begin{gather}
	\notag
	\braket{ \xi_a (t) \xi_a (t') + \xi_a (t') \xi_a (t) } = \gamma ( 2 n\s{th} + 1 ) \delta (t - t'),\ a=q,p;
	\\
	\notag
	\braket{ \xi_q (t) \xi_p (t') +  \xi_p (t') \xi_q (t) } = 0.
\end{gather}

The LN in adiabatic regime with intracavity modes eliminated is approximately given by (here $K_1 = K_2 = K$)
\begin{equation}
	\label{eq:adiab_bath}
	E_\eta \approx - \ln \frac{ 1 }{ 2 K^2 } \sqrt{ 1 + \Gamma K^4 + \frac{ K^2 }{ S^2 } ( 1 + \Gamma K^4 ) },
	\quad
	\Gamma = 2 \gamma \tau n\s{th}.
\end{equation}
In case of zero mechanical damping the expression is reduced to~\eqref{eq:symplectic}.

The LN corresponding to the full solution with all the imperfections is plotted as a function of the squeezing $S$ in Fig.~\ref{fig:loss_noise} for a set of different parameters.

The main means how the mechanical environment affects the entanglement is adding the thermal noise to the mechanical quadratures. Besides this the environment also creates small imbalance that prohibits the perfect cancellation of the optical mediator mode in $q_1$ by feedforward. The magnitude of this imbalance is  however almost negligible.

We as well plot the LN as a function of the squeezing for nonzero loss ($1 - \eta \neq 0$). The Fig.~\ref{fig:loss_noise} shows that at higher squeezing the entanglement between the mechanical modes is more tolerant to the mechanical bath than to the optical loss. Nevertheless, even with realistic loss parameters the entanglement does not vanish. We observe that adiabatic elimination is capable to very well fit the results for wide range of feasible squeezing of radiation.

Numerical analysis shows that the nonzero occupation of the mechanical bath creates a threshold for the coupling that allows the entanglement. At the same time, nonzero optical loss only decreases the value of the LN, so in case of zero occupation of the bath, the entanglement can tolerate any finite loss.

\subsection{Coupling optimization for experiments~\cite{meenehan_pulsed_2015,ockeloen-korppi_low-noise_2016}} 
\label{sec:coupling_optimization}

\begin{figure}[t]
	\centering
	\includegraphics[width=.95\linewidth]{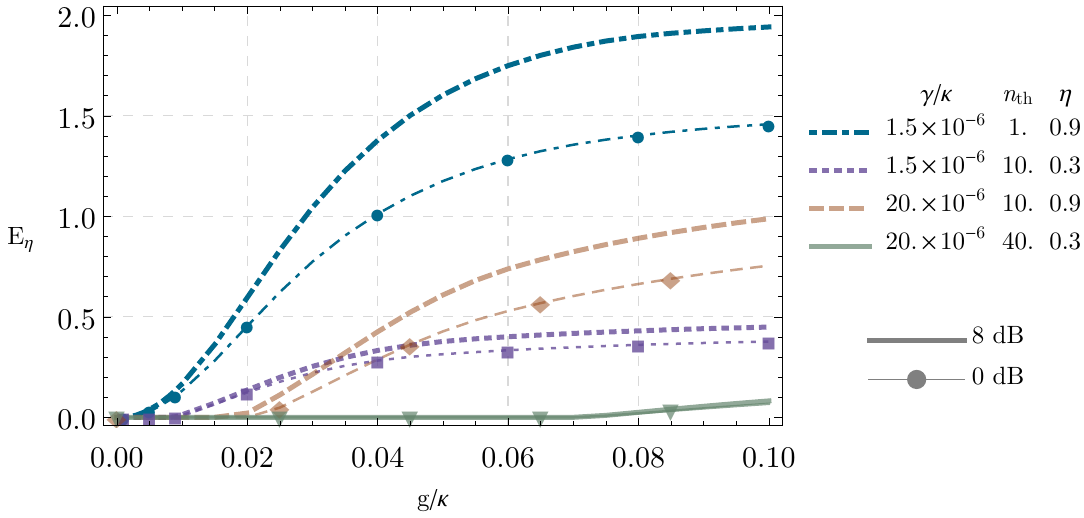}
	\caption{
		Maximal entanglement achievable with the coupling rate $g$ [in units of $\kappa$], for optomechanical parameters~\cite{meenehan_pulsed_2015} (blue dot-dashed and violet dotted lines) and electromechanical~\cite{ockeloen-korppi_low-noise_2016} (brown dashed and green solid).
	}
	\label{fig:optimized}
\end{figure}

In prior sections we focused on approaching a pure QND interaction between the two mechanical modes. Therefore we assumed the feedforward to be adjusted in a way that helps to cancel most of the optical mediator quadrature $X\up{in}$, i.e. $K_f = K_1 /\sqrt{ \eta}$.  Now we aim for maximization of the entanglement between the two modes. We waive the constraint on $K_f$ and numerically optimize the logarithmic negativity with respect to the optomechanical gains $K_{1,2}$, feedforward strength $K_f$, and the pulse duration $\tau$ given a limitation on the coupling strength.

The results of the numerical optimization are presented in Fig.~\ref{fig:optimized}. The optimal regime to achieve maximal entanglement appears to be very close to the regime of pure QND between the mechanical oscillators with long pulses $\kappa \tau \gg 1$ and $K_f = K_1/\sqrt{\eta}$.

Squeezed source of radiation apparently helps to improve entanglement in both opto- and electromechanical scheme for large $\eta$ close to perfect transmission and smaller $n\s{th}$. Simultaneously, the threshold for $g/\kappa$ to observe entanglement is lowered as well for higher $\eta$ and lower $n\s{th}$. On the other hand, for larger $n\s{th}$ and lower $\eta$, the squeezing of radiation is not important, however, we can still observe entanglement of mechanical systems if $\gamma/\kappa$ is not too large and $g/\kappa$ is sufficiently large.  Our analysis (see Appendix~\ref{sec:rotating_wave_approximation}) shows that under these conditions and for moderate squeezing the rotating wave approximation standardly employed in theory of optomechanics is well justified.  It is therefore fully feasible to generate entanglement with state-of-the-art systems.

The optomechanical setup noticeably outperforms the electromechanical one due to higher eigenfrequency of the mechanical oscillator and consequently lower bath occupation. The high occupation of the mechanical thermal bath in the electromechanical setup places a constraint on the available pulse durations which in turn limits the QND gain $K$.


\section{Conclusion}

We have proposed feasible way of the simplest pulsed implementation of entangling quantum non-demolition coupling between two distant but very similar mechanical oscillators, implementable with both current electromechanical and optomechanical setups. The method exploits squeezed light and microwave radiation and highly efficient homodyne detection to induce maximal entanglement for this purely mechanical coupling. We verified robustness of the procedure under small transmission loss between the oscillators and under mechanical thermal baths. We realized that both current optomechanical~\cite{meenehan_pulsed_2015} and electromechanical~\cite{ockeloen-korppi_low-noise_2016} setups are sufficient for the implementation of an extended version of multiple QND interaction. It will allow pulsed studies of quantum synchronization of mechanical objects~\cite{Mari2013,Ying2014,Weiss2016}. Afterwards, a detailed study of quantum interaction of possibly very different mechanical systems is important for development of physical connection with quantum thermodynamics~\cite{dong_work_2015,zhang_theory_2014,elouard_reversible_2015,brunelli_out--equilibrium_2015}. The method can be further extended to controllably couple more mechanical systems in future by different type of Gaussian interactions and possibly challenging non-Gaussian transformations.

\begin{acknowledgments}
	We acknowledge Project No. GB14-36681G of the
	Czech Science Foundation. A.A.R. acknowledges
	support by the Development Project of Faculty of Science,
	Palacky University. N.V. acknowledges the support
	of Palacky University (IGA-PrF-2016-005).
\end{acknowledgments}


\appendix

\section{Logarithmic negativity}
\label{LNapp}

The mechanical modes in our system are initially in thermal states and the optical modes are all in vacuum, and the linear dynamic preserves the Gaussianity of the states of mechanical modes.
A Gaussian state of a two-mode system with quadratures $f = [ q_1 , p_1 , q_2 , p_2 ]^T$ is fully determined by a vector of means $\avg f$ and a covariance matrix (CM) with elements defined as
\begin{equation}
	\notag
	V_{i j}
	= \frac 1 2 \braket{ \Delta f_i \Delta f_j + \Delta f_j \Delta f_i }.
\end{equation}
Here angular brackets denote the averaging over the quantum state, and $\Delta f_i \equiv f_i - \avg{ f_i }$.

Covariance matrix may be divided into $2 \times 2$ blocks such that:
\begin{equation*}
	 V =
	 \begin{bmatrix}
		\mathcal{V}_1 & \mathcal{V}_c \\
		\mathcal{V}_c^T & \mathcal{V}_2
	\end{bmatrix},
\end{equation*}
where $\mathcal{V}_1$ and $\mathcal{V}_2$ characterize internal correlations in mechanical subsystems. The matrix $\mathcal{V}_c$ stands for the correlations between the first and second mechanical modes. The diagonalisation of the CM leads to symplectic eigenvalues $\nu_{\pm}$~\cite{serafini_symplectic_2004}:
\begin{equation*}
	\nu_{\pm} = \sqrt{\frac{1}{2} \left(\Sigma(V) \pm \sqrt{\Sigma(V)^2 - 4 \det{V}} \right)},
\end{equation*}
with
\begin{equation*}
	\Sigma(V) = \det{\mathcal{V}_1} + \det{\mathcal{V}_2} - 2 \det{\mathcal{V}_c}.
\end{equation*}
Logarithmic negativity is defined then as $E_{{\eta}} = \max[0, -\ln{2 \nu_-}]$ and we use it as the measure of the entanglement of the system under the consideration.
\section{Beyond Rotating Wave Approximation} 
\label{sec:rotating_wave_approximation}

\begin{figure}[htb]
	\centering
	\includegraphics[width=.99\linewidth]{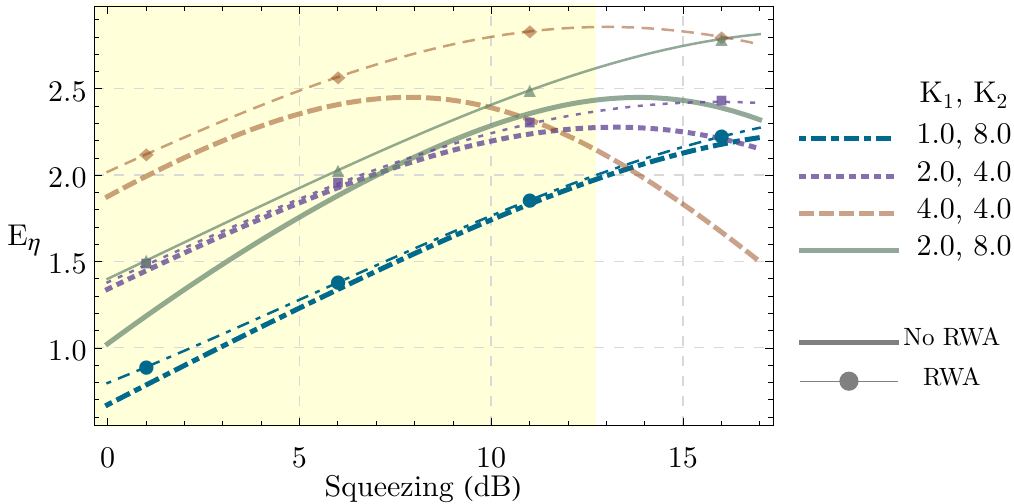}
	\caption{Entanglement (logarithmic negativity) as a function of squeezing of the input state computed from the full solution (thick lines) and with help of RWA (thin lines with markers).  For details see caption of~Fig.~\ref{fig:coupling}.}
	\label{fig:rwapic}
\end{figure}

The Rotating Wave Approximation is usually adopted for considerations of the optomechanical systems working in the resolved-sideband regime ($\kappa\ll\omega_m$).  In this Appendix we consider our protocol without this approximation.  We outline here the main steps that lead to an analytical expression for the covariance matrix of the mechanical modes.  The covariance matrix contains additional terms from back-action compared to the case of RWA.  We show that these terms do not impact the entanglement of the modes much.  For the sake of simplicity we do not consider in this Appendix thermal environments of mechanics and optical losses between the cavities.  Both these effects can be easily taken into account.

The equations of motion for the first system read
\begin{align}
	\label{eq:rwaq1}
	\dot q_1 & = g_1 X_1 (\cos 2 \omega_m t -1 ),
	\\
	\label{eq:rwap1}
	\dot p_1 & = g_1 X_1 \sin 2 \omega_m t,
	\\
	\label{eq:rwaX1}
	\dot X_1 & = \sqrt{2 \kappa } X_1^{\text{in}}-\kappa  X_1,
	\\
	\label{eq:rwaY1}
	\dot Y_1 & = \sqrt{2 \kappa } Y_1^{\text{in}}-\kappa  Y_1
	\\
	&
	\notag
	+ g_1 p_1  ( 1 - \cos 2 \omega_m t ) + g_1 q_1  \sin 2 \omega_m t.
\end{align}

As is easily seen, this system of equations allows an analytical solution.  First, the~Eq.~\eqref{eq:rwaX1} has the solution
\begin{equation}
	X_1 (t) = e^{ - \kappa t } \Big[ X_1 (0) + \sqrt{ 2 \kappa } \int_0^t ds \: e^{ \kappa s } X_1\up{in} (s) \Big].
\end{equation}
We then plug this expression into Eqs.~(\ref{eq:rwaq1},\ref{eq:rwap1}) to solve for $q_1$ and $p_1$.  The solution for $p_1$ reads
\begin{multline}
	\label{eq:rwap1sol}
	p_1 (\tau) - p_1 (0) = X_1 (0) g_1 \int_0^\tau dt \: e^{ - \kappa t } \sin 2 \omega_m t
	\\
	+ g_1 \sqrt{ 2 \kappa } \int_0^\tau dt \: e^{ - \kappa t } \sin 2 \omega_m t \int_0^t ds \: e^{\kappa s } X\up{in} (s)
	\\
	= X_1(0) g_1 \ci (0) + g_1 \sqrt{ 2 \kappa } \int_0^\tau ds e^{ \kappa s } X\up{in} (s) \ci (s ),
\end{multline}
where
\begin{equation}
	\ci (s ) \equiv  \int_s^\tau dt \: e^{ - \kappa t } \sin 2 \omega_m t.
\end{equation}
Notice that we swapped the order of integration when going to the last line in~\eqref{eq:rwap1sol} in order to have $X\up{in}$ in the outermost integration.  We do a similar swap with the consequent expressions.

The solution for $q_1$ can be written in a similar fashion.  This with~\eqref{eq:rwap1sol} can then be substituted into Eq.~\eqref{eq:rwaY1} to obtain the solution for $Y_1$.

The very same procedure repeatedly applied to the equations of motion for the second cavity and input-output relations allows to obtain a full analytical solution for the vector of quadratures of the mechanical modes.  The solution itself is rather cumbersome so we do not present it here.

Having the solution we proceed to compute the covariance matrix.  To demonstrate the method of calculation we use the~Eq.~\eqref{eq:rwap1sol} to compute the element~$V_{2,2} = \avg{ p_1 (\tau )^2 }$.
\begin{multline}
	\label{eq:vppsol}
	V_{2,2} = \avg{ p_1^2 (0) } + \avg{ X_1^2 (0) } g_1^2 \ci^2 (0)
	\\
	+ 2 \kappa g_1^2 \iint_0^\tau\!\!\!\! ds ds' \: \symm{ X_1\up{in} (s) }{ X_1\up{in} (s') } e^{ \kappa ( s + s' ) } \ci (s) \ci(s')
	\\
	= \avg{ p_1^2 (0) } + \avg{ X_1^2 (0) } g_1^2 \ci^2 (0)
	+ V\s{X} 2 \kappa g_1^2 \int_0^\tau \!\!\!\! ds \: e^{ 2 \kappa s } \ci^2 (s),
\end{multline}
where we used
\begin{equation}
	\symm{ X_1\up{in} (s) }{ X_1\up{in} (s') } = V\s{X} \delta ( s - s').
\end{equation}

It is illustrative to estimate the difference between the full solution~\eqref{eq:vppsol} and the straightforward solution $V_{2,2}\up{RWA} = \avg{p_1^2 (0)}$ obtained with advantage of RWA.  The quantity $\ci$ defined above serves as a measure of this divergence.  One can make estimations
\begin{align*}
	& (g_1 \ci (0) )^2 \sim \left( \frac{ g_1 }{ 2 \omega_m } \right)^2 =  \left( \frac{ g_1 }{ \kappa } \right)^2 \left( \frac{ \kappa }{ 2 \omega_m }\right)^2 \ll 1,
	\\
	& 2 \kappa g_1^2 \int_0^\tau ds \: e^{ 2 \kappa s } \ci^2 (s) \sim \cos^2 2 \omega_m \tau \left( \frac{ g_1 }{ 2 \omega_m }\right)^2 \ll 1.
\end{align*}

Besides this simple estimates we present the computed logarithmic negativity of the mechanical modes in Fig.~\ref{fig:rwapic}.  One can see that the adoption of RWA leads to an overestimation of entanglement due to the back-action that comes from the counterrotating terms in the Hamiltonian.  However, for appropriate parameters the full solution without RWA still approaches rather closely the idealized adiabatic one provided that the optomechanical coupling is not too strong (cf. blue dot-dashed lines in Fig.~\ref{fig:coupling} and~\ref{fig:rwapic}).  We use the sideband-resolution parameter $ \kappa / \omega_m  = 0.04$ which is a conservative estimate for a number of current experimental setups~\cite{Palomaki2013,meenehan_pulsed_2015}.

We became aware recently of another publication~\cite{malz_exact_2016} that deals with a QND interaction beyond RWA.

\bibliography{bib_two_qnd}
\end{document}